\documentclass[preprints,article,accept,moreauthors]{Definitions/mdpi}

\usepackage{xcolor}
\usepackage{subcaption}
\usepackage{ulem}

\definecolor{purple}{rgb}{0.62, 0.0, 0.77}

\firstpage{1} 
\pubvolume{1}
\issuenum{1}
\articlenumber{0}
\pubyear{}
\copyrightyear{}
\externaleditor{Academic Editor:}
\datereceived{} 
\daterevised{} 
\dateaccepted{} 
\datepublished{ } 
\hreflink{https://doi.org/} 

\Title{Chiral Invariant Mass Constraints from HESS J1731–347 in an Extended Parity Doublet Model with Isovector Scalar Meson}

\TitleCitation{Chiral Invariant Mass Constraints from HESS J1731–347 in an Extended Parity Doublet Model with the $a_0(980)$ Meson}

\Author{Yuk-Kei Kong$^{1,}$*$^{,\dagger}$, Bikai Gao $^{2,1}$*$^{,\dagger}$ and Masayasu Harada $^{3,1,4,}$*$^{,\dagger}$}

\AuthorNames{Yuk-Kei Kong, Bikai Gao and Masayasu Harada}

\AuthorCitation{Kong, Y.; 
 Gao, B.; \linebreak Harada, M.}

\address{%
$^{1}$ \quad Department of Physics, Nagoya University, Nagoya 464-8602, Japan \\
$^{2}$ \quad Research Center for Nuclear Physics (RCNP), Osaka University, Osaka 567-0047, Japan \\
$^{3}$ \quad Kobayashi-Maskawa Institute for the Origin of Particles and the Universe, Nagoya University, \linebreak Nagoya 464-8602, Japan \\
$^{4}$ \quad Advanced Science Research Center, Japan Atomic Energy Agency, Tokai 319-1195, Japan}

\corres{Correspondence: 
 yukkekong2-c@hken.phys.nagoya-u.ac.jp (Y.-K.K.); bikai@rcnp.osaka-u.ac.jp (B.G.); harada@hken.phys.nagoya-u.ac.jp (M.H.) 
 }

\firstnote{These authors contributed equally to this work.}

\abstract{
The recent discovery of a central compact object (CCO) within the supernova remnant HESS J1731-347, with mass $0.77^{+0.20}_{-0.17} \ M_\odot $ and radius $10.4^{+0.86}_{-0.78}$ km is the lightest and smallest compact object ever observed. We identify it as an ultra-light Neutron star (NS) and constrain the chiral invariant mass of nucleon $m_0$ from the observational data of NS using an extended parity doublet model with including the isovector scalar meson $a_0(980)$. We study the higher order asymmertic matter properties such as the symmetry incompressibility $K_{sym}$ and the symmetry skewness $Q_{sym}$ in the presence of $a_0$ meson. We find that $K_{sym}$ and $Q_{sym}$ is sensitive to the chiral invariant mass of nucleon $m_0$ in the presence of $a_0$ meson. We show that  the equation of state
in the present model satisfies
all observational constraints within $2\sigma$ credible region including the HESS J1731-347 observation, as well as the constraint from $K_{sym}$ when  
$740 \,\text{ MeV} \lesssim m_0 \lesssim 860 \,\text{ MeV}$ for $L_0 = $ 57.7 MeV. Yet, the $1\sigma$ constraint from neutron stars appears to be not
fully compatible with the constraint from $K_{sym}$ from the present model.
}

\keyword{parity doublet model; chiral invariant mass; isovector scalar meson; neutron star; HESS J1731-347 
}

\vspace{3pt}

\begin{document}

\section{Introduction}
Neutron star (NS) is one of the most compact objects in the universe. It is an excellent cosmic laboratory under extreme conditions for studying dense QCD matter. NSs allow us to study the equation of state (EoS) 
of the QCD matter in high density, which is difficult to access in the experiments. Recently, more and more NS observations are available and provides us valuable information about the EoS. For example, the NS merger event GW170817 provided insights into the mass and radius of NSs, with an estimation of approximately 1.4 $M_\odot$ and a radius of $R = 11.9^{+1.4}_{-1.4}$ km~\cite{LIGOScientific:2017vwq, LIGOScientific:2018cki}. The NS observations from NICER has also played a crucial role in advancing our understandings of NSs. The analyses~\cite{Miller_2021,Riley:2021pdl} have focused on NSs with masses around 1.4 $M_\odot$ and $2.1M_\odot$. The results show that the radii of these NSs are rather similar for different masses, with a radius of approximately 12.45$\pm$0.65 kilometers for a 1.4 $M_\odot$ NS and 12.35$\pm$0.75 kilometers for a
2.08 $M_\odot$ NS.

Recently, report on a central compact object (CCO) HESS J1731-347~\cite{HESSJ1731-347} 
with very small mass $M = 0.77^{+0.20}_{-0.17}$ $M_\odot$ and radius $R = 10.4^{+0.86}_{-0.78}$ km has challenging our understanding to the NS. The observation of HESS J1731-347 implies that the NS EoS
is very soft in the low-density region. There are also studies that consider HESS J1731-347 as a quark star~\cite{Chu:2023rty,Oikonomou:2023otn,Yang:2023haz,Kourmpetis:2025zjz}, 
an exotic object made from 
deconfined quarks rather than the usual hadronic matter. Understanding this CCO is therefore important to the study of NS and the EoS.

Chiral symmetry and its spontaneous breaking play a fundamental role in quantum chromodynamics (QCD) and low-energy hadron physics. This symmetry breaking is responsible for the generation of the hadron masses and the mass differences between chiral partners. In dense environments, such as the interior of NSs, chiral symmetry is expected to be partially restored. Investigating how hadron masses change under such conditions can provide valuable insights into the origin of hadron masses and the properties of strongly interacting matter.

The Parity Doublet Model (PDM)~\cite{PhysRevD.39.2805} is an extended linear sigma model that incorporates a parity doubling structure of nucleons. In this model, the negative-parity excited nucleon is considered as the chiral partners of the ground state nucleons, with spontaneous symmetry breaking generating the mass difference between them. When chiral symmetry is restored, these nucleons degenerate into the same mass called chiral invariant mass of nucleon $m_0$. Studies such as lattice simulations ~\cite{PhysRevD.92.014503,2017JHEP...06..034A} and QCD sum rule~\cite{PhysRevD.105.014014} suggest that part of the nucleon mass is independent of the chiral symmetry breaking. Both quantitative and qualitative study of the chiral invariant mass are therefore crucial for advancing our understanding of the origin of hadron masses.

Previous analyses have attempted to constrain $m_0$ by analyzing nucleon properties in vacuum. Ref.~\cite{10.1143/PTP.106.873} suggests that $m_0$ is smaller than 500 MeV based on an analysis of the decay width of $N(1535)$, while Ref.~\cite{Yamazaki:2018stk} includes higher-derivative interactions in the model, resulting in a larger values of $m_0$ that consistent with the decay width from experiments.

The PDM has also been applied to study dense medium in several studies, such as in 
Refs.~\cite{Hatsuda:1988mv, Zschiesche:2006zj, Dexheimer:2007tn, Dexheimer:2008cv, Sasaki:2010bp, Sasaki:2011ff,%
Gallas:2011qp, Paeng:2011hy,%
Steinheimer:2011ea,Dexheimer:2012eu, Paeng:2013xya,Benic:2015pia,Motohiro:2015taa,%
Mukherjee:2016nhb,Suenaga:2017wbb,Takeda:2017mrm,Mukherjee:2017jzi,Paeng:2017qvp,%
Marczenko:2017huu,Abuki:2018ijb,Marczenko:2018jui,Marczenko:2019trv,Yamazaki:2019tuo,%
Harada:2019oaq,Marczenko:2020jma,Harada:2020etl,%
PhysRevC.103.045205,Marczenko:2021uaj,PhysRevC.104.065201,Marczenko:2022hyt,%
PhysRevC.106.065205,Minamikawa:2023eky,Minamikawa:2023ypn,Gao:2024mew,Marczenko:2023ohi,PhysRevC.109.065807,Gao:2024lzu,Gao:2024jlp,Yuan:2025dft,Gao:2025ncu}.
Recently, several studies~\cite{PhysRevC.103.045205,PhysRevC.104.065201,PhysRevC.106.065205,Minamikawa:2023eky,PhysRevC.109.065807} have constructed the NS EoS using an extended PDM~\cite{Motohiro:2015taa}. In these studies, the hadronic EoS is smoothly interpolated to a NJL-type quark matter EoS, under the assumption of a crossover hadron–quark phase transition, following the approach of Refs.~\cite{Baym_2018,Baym_2019}. 
Reference~\cite{PhysRevC.103.045205}
had constrained the chiral invariant mass of nucleon to $600$ MeV $\lesssim m_0 \lesssim 900$ MeV using the observational data of NS given in 
Refs.~\cite{NANOGrav:2019jur,LIGOScientific:2017vwq,LIGOScientific:2017ync, 
LIGOScientific:2018cki,Miller:2019cac,Riley:2019yda}. The constraint was updated to $400$ MeV$\lesssim m_0 \lesssim 700$ MeV by considering the effect of anomaly~\cite{PhysRevC.106.065205,Minamikawa:2023eky} with new NS data 
analysis~\cite{Fonseca:2021wxt,De:2018uhw,Radice:2017lry}. Reference~\cite{PhysRevC.108.055206}
constrained $m_0$ to $580$ MeV$\lesssim m_0 \lesssim 860$ MeV with the presence of isovector scalar meson. Reference~\cite{PhysRevC.109.065807}
showed that $m_0 \simeq 850$ MeV with the consideration of central compact object (CCO) within the supernova remnant HESS J1731-347~\cite{HESSJ1731-347}. The discovery of this ultra-light compact object HESS J1731-347, with a mass of approximately $0.77^{+0.20}_{-0.17} \, M_{\odot}$ and a radius of about $10.4^{+0.86}_{-0.78}$ km, has opened a new window for studying compact objects and provides additional constraints on the dense matter equation of state. As the lightest and smallest compact object ever observed, it challenges existing theories and requires careful theoretical modeling.

Recently, the effect of isovector-scalar $a_0$(980) meson (also called the $\delta$ meson) on asymmetric matter such as NS is rising attention. The $a_0$(980) meson accounts for the attractive force in the isovector channel. The effect of $a_0$(980) to the symmetry energy and asymmetric matter EoS was studied in Refs.~\cite{Kubis_1997,https://doi.org/10.48550/arxiv.astro-ph/9802303,Miyatsu_2022,Li_2022,https://doi.org/10.48550/arxiv.2209.02861,Thakur_2022,Liu_2005,PhysRevC.80.025806,Gaitanos_2004,PhysRevC.67.015203,PhysRevC.65.045201} using Walecka-type relativistic mean-field (RMF) models, and Refs.~\cite{PhysRevC.90.055801,PhysRevC.84.054309} using density-dependent RMF models. 
The existence of $a_0$(980) is shown to increase
the symmetry energy~\cite{Kubis_1997,Miyatsu_2022,Li_2022,Liu_2005,PhysRevC.80.025806,Gaitanos_2004,PhysRevC.67.015203,PhysRevC.65.045201}, and stiffen
the NS EoS~\cite{https://doi.org/10.48550/arxiv.astro-ph/9802303,Liu_2005,Li_2022,Miyatsu_2022,Thakur_2022} and asymmetric matter EoS~\cite{PhysRevC.84.054309}. 
Recently, the stiffening effect of $a_0(980)$ on the NS EoS was also confirmed
in an extended PDM, and  
the constraint to the chiral invariant mass is obtained as
$580$ MeV $\lesssim m_0 \lesssim 860$ MeV in Ref.~\cite{PhysRevC.108.055206}. The stiffening of the NS EoS due to the $a_0$ meson therefore may make
the interpretation of HESS 1731-347 as an ultra-light NS difficult.

In this work, we extend previous studies by incorporating the isovector scalar meson $a_0(980)$ into the PDM and explore its effects on the properties of neutron stars, with updated NS observations including the ultra-light compact object HESS J1731-347. Through this analysis, we aim to provide tighter constraints on the chiral invariant mass $m_0$ with the consideration of HESS J1731-347.

The paper is organized as follows: in Section~\ref{sec:PDMconstr}, we review an extension of the PDM by including the isovector scalar meson $a_0(980)$ based on the chiral U(2)$_L$ $\times$ U(2)$_R$ symmetry with U(1)$_A$ anomaly included, constructed in Ref.~\cite{PhysRevC.108.055206}. In Section~\ref{sec:PDMmatter} we construct the matter with PDM under the mean field approximation. 
Then, in Section~\ref{sec:a0}, 
we study effect of $a_0$ meson and chiral invariant mass of nucleon $m_0$ to the asymmetric matter properties such as the symmetry incompressibility $K_{sym}$ and the symmetry skewness coefficient $Q_{sym}$. By comparing to the results to recent $K_{sym}$ constraints from experiments and theoretical works, we constrain the value of $m_0$. In Section.~\ref{sec:NS}, we study neutron star matter using a unified equation of state with hadron-quark crossover, and analyze the mass-radius relationship to constrain the model parameters $m_0$ and $L_0$ using recent NS observations, including the ultra-light compact object HESS J1731-347. We then compare the constraint from nuclear matter properties to those constraint from NS observations. Finally, a summary is given in Section~\ref{sec:summary}.

\section{Dense Nuclear Matter with Parity Doublet Model}
\label{sec:PDM MF}

\subsection{A Parity Doublet Model with U(2)$_L \times$U(2)$_R$ Symmetry}
\label{sec:PDMconstr}

In this work, we use an parity doublet model (PDM) based on the U(2)$_L$ $\times$ U(2)$_R$ chiral symmetry constructed in Ref.~\cite{PhysRevC.108.055206}. The Lagrangian is given by 
\begin{equation}
  \mathcal{L} = \mathcal{L}_N + \mathcal{L}_M + \mathcal{L}_{V}\ , \label{PDM-L1}
\end{equation} 
where $\mathcal{L}_N$ is the nucleons, $\mathcal{L}_M$ the scalar and pseudoscalar mesons and $\mathcal{L}_{V}$ the vector mesons Lagrangian.

In this model, the scalar meson field $M$ is introduced as the $(2,2)_{-2}$ representation under the SU(2)$_L\times$SU(2)$_R\times$U(1)$_A$ symmetry, which transforms as 
\begin{equation}
  M \rightarrow e^{-2i\theta_A} g_L M g_R^{\dagger}\ ,
\end{equation} 
where $g_{R,L} \in \mbox{SU}(2)_{R,L} $ and $e^{-2i\theta_A} \in \mbox{U}(1)_{A} $. 
$M$ is parameterized as 
\begin{equation}
  M = [\sigma + i\vec{\pi} \cdot \vec{\tau}] -  [ \vec{a_0} \cdot \vec{\tau} + i\eta]\ ,
  \label{Mmotohiro}
\end{equation} 
where $\sigma, \vec{\pi}, \vec{a_0}, \eta$ are the sigma meson, pions, the lightest isovector scalar meson $a_0(980)$ and $\eta$ meson field, respectively. $\vec{\tau}$ are the Pauli matrices. The vacuum expectation value (VEV) of $M$ is given by
\begin{equation}
\begin{aligned}
  \langle 0 | M | 0 \rangle = \begin{pmatrix}
\sigma_0 & 0\\
0 & \sigma_0
\end{pmatrix},
\end{aligned}
\label{def M 2}
\end{equation} 
where $\sigma_0 = \langle 0|\sigma|0 \rangle$ is the VEV of the $\sigma$ field equal to the pion decay constant $f_\pi = 93 $ MeV.
Then, the Lagrangian $\mathcal{L}_M$ is given by
\begin{equation}
\mathcal{L}_M = \frac{1}{4} \mbox{tr}\left[\partial_\mu M \partial^\mu M^\dagger \right] - V_M\ ,
\end{equation} 
where $V_M$ is the potential for $M$. In the current model, $V_M$ is given by~\cite{PhysRevC.108.055206}
\begin{align}
V_M=&-\frac{\bar{\mu}^2}{4}\mbox{tr}[M^\dagger M ]+\frac{\lambda_{41}}{8}\mbox{tr}[(M^\dagger M)^2] \nonumber \\ 
&{}-\frac{\lambda_{42}}{16}\{ \mbox{tr}[M^\dagger M]\}^2-\frac{\lambda_{61}}{12}\mbox{tr}[(M^\dagger M)^3] \nonumber \\ 
&{}-\frac{\lambda_{62}}{24}\mbox{tr}[(M^\dagger M)^2]\mbox{tr}[M^\dagger M]-\frac{\lambda_{63}}{48}\{ \mbox{tr}[M^\dagger M]\}^3 \nonumber \\
&{}-\frac{m^2_{\pi}f_{\pi}}{4}\mbox{tr}[M+M^\dagger]-\frac{K}{8}\{\det M+\det M^\dagger\} \ ,
\label{VM}
\end{align} 
where terms up to the sixth order that are invariant under SU(2)$_L$ $\times$ SU(2)$_R\times$U(1)$_A$ symmetry are included. In addition, a determinant-type Kobayashi–Maskawa–’t Hooft interaction is included in the current model to implement the U(1)$_A$ anomaly.

The iso-triplet $\rho$ meson and iso-singlet $\omega$ meson are considered based on the hidden local symmetry (HLS)~\cite{PhysRevLett.54.1215,Bando:1987br,Harada:2003jx}. The HLS is introduced by performing polar decomposition of the field $M$ as
\begin{equation}
\begin{aligned}
M = \xi_L^\dagger S \xi_R \, ,
\end{aligned}
\label{HLSM}
\end{equation}
where $S = \sigma + \sum\limits_{b=1}^3 {a}_0^b \tau_b/2$ is the $2\times2$ matrix field for scalar mesons. $\xi_{L,R}$ are transforming as 
\begin{equation}
\begin{aligned}
\xi_{L,R} \rightarrow h_\omega h_\rho \xi_{L,R} g_{L,R}^\dagger e^{\pm i \theta_A},
\end{aligned}
\label{HLSXItr}
\end{equation}
where $h_\omega \in \mbox{U(1)}_{\rm HLS}$ and $h_\rho \in \mbox{SU(2)}_{\rm HLS}$. We note that $e^{+ i \theta_A}$ for $\xi_{L}$ and $e^{- i \theta_A}$ for $\xi_{R}$.
In the unitary gauge of the HLS, $\xi_{L,R}$ are parameterized as
\begin{equation}
\xi_R = \xi_L^\dagger 
= \exp \left( i P / f_\pi\right) \ ,
\label{HLSXI}
\end{equation}
where $P = \eta +  \sum\limits_{a=1}^{3}  \pi^a \tau_a/2 $ is the $2\times2$ matrix field for pseudoscalar mesons. The vector mesons are the gauge bosons in HLS and transform as 
\begin{equation}
\begin{aligned}
\omega_\mu \rightarrow h_\omega \omega_\mu h_\omega^\dagger + \frac{i}{g_\omega}\partial_\mu h_\omega h_\omega^\dagger ,
\end{aligned}
\label{HLSomega}
\end{equation}
\begin{equation}
\begin{aligned}
\rho_\mu \rightarrow h_\rho \rho_\mu h_\rho^\dagger + \frac{i}{g_\rho}\partial_\mu h_\rho h_\rho^\dagger ,
\end{aligned}
\label{HLSrho}
\end{equation}
where $\omega_\mu$ and $\rho_\mu = \sum\limits_{a=1}^3 \rho_\mu^a \tau_a/2$ are the gauge bosons of SU(2)$_{\rm HLS}$ and U(1)$_{\rm HLS}$, respectively. $g_\omega$ and $g_\rho$ are the corresponding HLS gauge coupling constants.

The HLS-invariant Lagrangian is given by

\begin{equation}
\begin{aligned}
\mathcal{L}_{V} = {} & a_{VNN} \left[ \Bar{N}_{1l} \gamma^\mu \xi_L^\dagger \hat{\alpha}_{\parallel \mu} \xi_L N_{1l} + \Bar{N}_{1r} \gamma^\mu \xi_R^\dagger \hat{\alpha}_{\parallel \mu} \xi_R N_{1r} \right] \\ 
  & + a_{VNN} \left[ \Bar{N}_{2l} \gamma^\mu \xi_R^\dagger \hat{\alpha}_{\parallel \mu} \xi_R N_{2l} + \Bar{N}_{2r} \gamma^\mu \xi_L^\dagger \hat{\alpha}_{\parallel \mu} \xi_L N_{2r} \right] \\ 
  & + a_{0NN} \sum_{i = 1,2} \left[ \Bar{N}_{il} \gamma^\mu \mbox{tr} [ \hat{\alpha}_{\parallel \mu} ] N_{il} + \Bar{N}_{ir} \gamma^\mu \mbox{tr} [\hat{\alpha}_{\parallel \mu} ] N_{ir} \right] \\
& + \frac{{m_\rho}^2}{{g_\rho}^2}\mbox{tr} [\hat{\alpha}_\parallel^\mu \hat{\alpha}_{\parallel\mu}] + \left(\frac{{m_\omega}^2}{8{g_\omega}^2} - \frac{{m_\rho}^2}{2{g_\rho}^2}\right) \mbox{tr} [\hat{\alpha}_\parallel^\mu]\mbox{tr}[ \hat{\alpha}_{\parallel\mu}] - \frac{1}{8{g_\omega}^2}\mbox{tr}[\omega^{{\mu\nu}}\omega_{{\mu\nu}}] - \frac{1}{2{g_\rho}^2}\mbox{tr}[\rho^{{\mu\nu}}\rho_{{\mu\nu}}] \\
& + \lambda_{\omega \rho}\left( a_{VNN} + a_{0NN} \right)^2 a_{VNN}^2 \left[ \frac{1}{2}\mbox{tr} [\hat{\alpha}_\parallel^\mu \hat{\alpha}_{\parallel\mu}]\mbox{tr} [\hat{\alpha}_\parallel^\nu]\mbox{tr}[ \hat{\alpha}_{\parallel\nu}] - \frac{1}{4}\left\{ \mbox{tr} [\hat{\alpha}_\parallel^\mu]\mbox{tr}[ \hat{\alpha}_{\parallel\mu}] \right\}^2 \right] ,
\end{aligned}
\label{Lv}
\end{equation}

where $\rho^{{\mu\nu}}$ and $\omega^{{\mu\nu}}$ are the field strengths of $\rho$ meson and $\omega$ meson that given by
\begin{align}
\rho_{\mu\nu} = & \partial_\mu \rho_\nu - \partial_\nu \rho_\mu - i g_\rho \, \left[ \rho_\mu \,,\, \rho_\nu \right] \ , \notag\\
\omega_{\mu\nu} = & \partial_\mu \omega_\nu - \partial_\nu \omega_\mu \ .
\end{align}
$\hat{\alpha}_\perp^\mu$ and $\hat{\alpha}_\parallel^\mu$ are the covariantized Maurer$-$Cartan 1-forms defined as
\begin{equation}
  \hat{\alpha}_\perp^\mu \equiv \frac{1}{2i}[D^\mu \xi_R \xi_R^\dagger - D^\mu \xi_L \xi_L^\dagger] ,
\end{equation}
\begin{equation}
  \hat{\alpha}_\parallel^\mu \equiv \frac{1}{2i}[D^\mu \xi_R \xi_R^\dagger + D^\mu \xi_L \xi_L^\dagger] ,
\end{equation}
and the covariant derivatives of $\xi_{L,R}$ are given by 
\begin{align}
  D^\mu \xi_L = \partial^\mu \xi_L - ig_\rho \rho^\mu \xi_L - ig_\omega \omega^\mu \xi_L + i \xi_L \mathcal{L}^\mu - i \xi_L \mathcal{A}^\mu \ ,
\\
  D^\mu \xi_R = \partial^\mu \xi_R - ig_\rho \rho^\mu \xi_R - ig_\omega \omega^\mu \xi_R + i \xi_R \mathcal{R}^\mu + i \xi_R \mathcal{A}^\mu \ ,
\end{align}
with $\mathcal{L}^\mu$, $\mathcal{R}^\mu$ and $\mathcal{A}^\mu$ being the external gauge fields corresponding to SU(2)$_L \times$SU(2)$_R \times$U(1)$_A$ global symmetry. 

The last term in the Lagrangian~(\ref{Lv}) is a mixing interaction of $\rho$ and $\omega$ mesons as introduced in Ref.~\cite{PhysRevC.108.055206} to reduce the slope parameter, following Ref.~\cite{PhysRevC.106.065205}.

The baryonic Lagrangian $\mathcal{L}_N$ based on the parity doubling \cite{PhysRevD.39.2805,10.1143/PTP.106.873} is given by 
\begin{equation}
\begin{aligned}
\mathcal{L_N} {} & = \Bar{N_{1}}i \gamma^\mu \mathcal{D}_\mu N_{1} + \Bar{N_{2}}i \gamma^\mu \mathcal{D}_\mu N_{2} \\
 & \quad - m_0 [\Bar{N}_{1}\gamma_5 N_{2} - \Bar{N}_{2}\gamma_5 N_{1}] \\
   & \quad - g_1 [\Bar{N}_{1l} M N_{1r} + \Bar{N}_{1r} M^{\dagger} N_{1l}]\\
    & \quad - g_2 [\Bar{N}_{2r} M N_{2l} + \Bar{N}_{2l} M^{\dagger} N_{2r}], \
\end{aligned}
\label{eq231}
\end{equation}
where $N_{ir} = \frac{1+\gamma_5}{2} N_i$ ($N_{il}= \frac{1-\gamma_5}{2}N_i$) ($i=1,2$) is the left-handed (right-handed) component of the nucleon fields $N_i$ and the covariant derivatives are defined as
\begin{equation}
  \begin{aligned}
    \mathcal{D}^\mu N_{1l,2r} & = \left( \partial^\mu - i \mathcal{L}^\mu - i {\mathcal V}^\mu + i {\mathcal A}^\mu \right) N_{1l,2r}\ ,\\
    \mathcal{D}^\mu N_{1r,2l} & = \left( \partial^\mu - i \mathcal{R}^\mu - i {\mathcal V}^\mu - i {\mathcal A}^\mu \right) N_{1r,2l}\ ,
  \end{aligned}
\label{eq231}
\end{equation}
where ${\mathcal V}^\mu$ is the external gauge field corresponding to the U(1) baryon number. $g_1$ and $g_2$ are the Yukawa couplings of the nucleon $N_i$ and $m_0$ is called the chiral invariant mass of nucleon. Two baryon fields $N_+$ and $N_-$ corresponding to the positive parity and negative parity nucleon fields can be obtained by diagonalizing $\mathcal{L}_N$ and their vacuum masses are given by~\cite{PhysRevD.39.2805,10.1143/PTP.106.873}
\begin{equation}
\begin{aligned}
  m^{\rm(vac)}_{\pm} = \frac{1}{2} \bigg[ \sqrt{(g_1+g_2)^2\sigma_0^2 + 4m_0^2} \pm (g_1 - g_2)\sigma_0 \bigg]\ .
\end{aligned}
\label{mvaj}
\end{equation} 


In the present work, we identify the fields $N_+$ and $N_-$ as the ground state $N$(939) and its excited state $N(1535)$.

\subsection{PDM with Mean Field Approximation}\label{sec:PDMmatter}

In this work, the mean-field approximation is adopted as in~\cite{PhysRevC.108.055206}
\begin{equation}
  \sigma(x) \rightarrow \sigma, \qquad \pi(x) \rightarrow 0, \qquad a_0^i(x) \rightarrow a\,\delta_{i3}, \qquad \eta(x) \rightarrow 0,
\end{equation} 
and the matrix $M$ is given by
\begin{equation}
\left\langle M \right\rangle = \begin{pmatrix}
\sigma - a & 0 \\ 0 & \sigma + a
\end{pmatrix}
\ .
\end{equation}
Then, the mean potential $V_M$ is written as
\begin{equation}
\begin{aligned}
V_M = & - \frac{\bar{\mu}^2_{\sigma}}{2} \sigma^2 - \frac{\bar{\mu}^2_{a}}{2} a^2 + \frac{\lambda_4}{4} (\sigma^4 + a^4 ) + \frac{\gamma_4}{2} \sigma^2 a^2 \\ 
& {} - \frac{\lambda_6}{6} (\sigma^6 +15\sigma^2a^4 + 15\sigma^4a^2 +a^6 ) +\lambda_6^{'}(\sigma^2a^4 + \sigma^4a^2) \\ 
& {} - m^2_{\pi} f_{\pi} \sigma \ ,
\end{aligned}
\label{VMa}
\end{equation}
where the parameters are redefined as 
\begin{equation}
\begin{aligned}
{} & \bar{\mu}_{\sigma}^2 \equiv \bar{\mu}^2 + \frac{1}{2} K\ , \\
& \bar{\mu}^2_a \equiv \bar{\mu}^2 - \frac{1}{2} K = \bar{\mu}_{\sigma}^2 - K\ , \\
& \lambda_4 \equiv \lambda_{41} - \lambda_{42}\ , \\
& \gamma_4 \equiv 3\lambda_{41} - \lambda_{42}\ , \\
& \lambda_6 \equiv \lambda_{61} + \lambda_{62} + \lambda_{63}\ , \\
& \lambda_6^{'} \equiv \frac{4}{3}\lambda_{62} + 2\lambda_{63}\ . 
\label{eq2.23}
\end{aligned}
\end{equation}

The vector meson mean fields are given by
\begin{equation}
  \omega_{\mu} (x) \rightarrow \omega \delta_{\mu 0}, \qquad  \rho^{i}_{\mu} (x) \rightarrow \rho \delta_{\mu 0} \delta_{i 3},
\end{equation}
and the mean field Lagrangian of the vector mesons is given by
\begin{equation}
\begin{aligned}
\mathcal{L}_{V} = & - g_{\omega NN} \sum_{\alpha j} \Bar{N}_{\alpha j} \gamma^0 \omega N_{\alpha j} - g_{\rho NN} \sum_{\alpha j} \Bar{N}_{\alpha j} \gamma^0 \frac{\tau_3}{2} \rho N_{\alpha j} \\
& + \frac{1}{2} m^2_{\omega} \omega^2 + \frac{1}{2} m^2_{\rho} \rho^2 + \lambda_{\omega \rho} g_{\omega NN}^2 g_{\rho NN}^2 \omega^2 \rho^2\ . 
\end{aligned}
\end{equation} 
with 
\begin{align}
  g_{\omega NN} & = \left( a_{VNN} + a_{0NN} \right) g_\omega \ ,\\
  g_{\rho NN} & = a_{VNN}g_\rho \ .
\end{align}

The thermodynamic potential of the nucleons is given by 
\begin{equation}
\begin{aligned}
{} & \Omega_{N}  = - 2 \sum_{\alpha=\pm, j=\pm} \int^{k_f} \frac{d^3p}{(2 \pi)^3} \bigg[ \mu^*_j - \omega_{\alpha j} \bigg],
\end{aligned}
\label{OFG}
\end{equation} 
$\alpha = \pm$ denotes the parity and $j = \pm $ the iso-spin of nucleons ($j=+$ for proton and $j=-$ for neutron).
The effective chemical potential $\mu^*_j$ is given by 
\begin{equation}
\begin{aligned}
\mu^*_j \equiv (\mu_B - g_{\omega NN} \omega) + \frac{j}{2} (\mu_I - g_{\rho NN} \rho)\ .
\end{aligned}
\label{mustar}
\end{equation} 
$\omega_{\alpha j}$ is the nucleon energy 
as defined by $\omega_{\alpha j} = \sqrt{(\vec{p})^2 + (m^*_{\alpha j})^2}$, where $\vec{p}$ and $m^*_{\alpha j}$ are the momentum and the effective mass of the nucleon. In the present model, the effective mass $m^*_{\alpha j}$ is given by 
\begin{equation}
\begin{aligned}
  m^*_{\alpha j} = \frac{1}{2} \bigg[ \sqrt{(g_1+g_2)^2(\sigma - ja)^2 + 4m_0^2} + \alpha(g_1 - g_2)(\sigma - ja) \bigg]\ .
\end{aligned}
\label{maj}
\end{equation}

Altogether, the hadronic thermodynamic potential is 
\begin{equation}
\begin{aligned}
{} & \Omega_H  = \Omega_{N} \\
& \qquad \;\; - \frac{\bar{\mu}^2_{\sigma}}{2} \sigma^2 - \frac{\bar{\mu}^2_{a}}{2} a^2 + \frac{\lambda_4}{4} (\sigma^4 + a^4 ) + \frac{\gamma_4}{2} \sigma^2 a^2 \\
& \qquad \;\; - \frac{\lambda_6}{6} (\sigma^6 +15\sigma^2a^4 + 15\sigma^4a^2+ a^6 )  + \lambda_6^{'}(\sigma^2a^4 + \sigma^4a^2) \\ 
& \qquad \;\; - m^2_{\pi} f_{\pi} \sigma - \frac{1}{2} m^2_{\omega} \omega^2 - \frac{1}{2} m^2_{\rho} \rho^2 - \lambda_{\omega \rho} g_{\omega NN}^2 g_{\rho NN}^2 \omega^2 \rho^2\\
& \qquad \;\; - \Omega_{0}\ ,
\end{aligned}
\label{eq36m}
\end{equation} 
where the vacuum potential
\begin{equation}
\begin{aligned}
\Omega_0 \equiv - \frac{\bar{\mu}^2_{\sigma}}{2} f_{\pi}^2 + \frac{\lambda_4}{4} f_{\pi}^4 
 - \frac{\lambda_6}{6} f_{\pi}^6 - m^2_{\pi} f_{\pi}^2\ .
\end{aligned}
\label{eq37}
\end{equation}      
is subtracted from $\Omega_H$.

\subsection{Determination of Model Parameters}\label{sec:DMP1}

In the present model, the model parameters are determined to reproduce the nuclear saturation properties and the vacuum properties of the hadrons. There are 11 parameters to be determined for a given value of chiral invariant mass $m_0$: 
\begin{equation}
  g_1\,,\ g_2\,,\ \bar{\mu}^2_{\sigma}\,, \ \bar{\mu}^2_{a} \,, \ \lambda_{4}\, ,\ \gamma_{4}\, ,\ \lambda_{6}\, ,\ \lambda'_{6}\,,\ g_{\omega NN}\,,\  g_{\rho NN}\, ,\  \lambda_{\omega \rho}\,.
\end{equation}
The vacuum expectation value of $\sigma$ is taken to be $\sigma_0=f_{\pi}$ with the pion decay constant $f_{\pi}=$ 92.4 MeV. The Yukawa coupling constants
$g_1$ and $g_2$ are determined by fitting to the nucleon masses in vacuum given in Equation~(\ref{mvaj}). 
In this study, we identify the nucleon as $N(939)$ and its parity partner as the excited state $N^*(1535)$ with $m_+ = m_{N}=939$ MeV and $m_{-} = m_{N^*}=1535$ MeV.
The values of 
$\bar{\mu}^2_{\sigma}$, $\lambda_4$, $\lambda_6$, $g_{\omega NN}$ are determined from the saturation properties: saturation density $n_0$, the binding energy $B_0$, and the incompressibility $K_0$ together with the stationary condition of the potential in vacuum,
\begin{equation}
\bar{\mu}_\sigma^2 f_\pi - \lambda_4 f_\pi^3 + \lambda_6 f_\pi^5 + m_\pi^2 f_\pi = 0 \ .
\label{vacuum condition}
\end{equation}
The value of the nuclear saturation properties are summarized in Table~\ref{SP}.
As investigated in Ref.~\cite{PhysRevC.108.055206} and Ref.~\cite{sym160912382024}, terms with coefficient $\lambda'_6$ are of sub-leading order in the large $N_c$ expansion and have small effect to the matter properties. Therefore, we set $ \lambda'_6 = 0 $ in this work for simplicity.
The parameters $\bar{\mu}^2_{a} = \bar{\mu}^2_{\sigma} - K$ and $\gamma_4$ are fitted to the meson masses and the other parameters
\begin{align}
K = & \quad m_\eta^2 - m_\pi^2 \ , \notag\\
\gamma_4 = & \quad \frac{m_{a_0}^2 + (5 \lambda_{6} - 2\lambda_{6}') f_{\pi}^4 + \bar{\mu}^2_{a}}{f_{\pi}^2} \ ,
\end{align}
where $m_\eta$ and $m_{a_0}$ are the masses of $\eta$ and $a_0(980)$. The values of meson masses used in this work are listed in Table~\ref{vacM}. The values of the parameters for various $m_0$ are presented in Table~\ref{PD1215}.
In the present model, the vector mixing interaction with coefficient $\lambda_{\omega\rho}$ are included to control behavior of the asymmetric matter at density $n_B > n_0$ beyond the saturation . The parameters $g_{\rho NN}$ and $\lambda_{\omega\rho}$ are related and fitted to the symmetry energy $S_0$ as well as the slope parameter $L_0$. As summarized in Ref.~\cite{universe7060182}, the recent accepted value of $L_0 = 57.7\pm 19$ MeV. Therefore, we carry out the calculations over the range $L_0 = 40$--$80$ MeV in this work.  The values of $g_{\rho NN}$ and $\lambda_{\omega\rho}$ are shown in Tables~\ref{tab:gr_lor_215} and \ref{tab:lwr_lor_215}.

\begin{table}[H] 
\caption{Saturation properties that are used to determine the model parameters: saturation density $n_0$, binding energy $B_0$, incompressibility $K_0$, and symmetry energy $S_0$.\label{SP}}
\newcolumntype{C}{>{\centering\arraybackslash}X}
\begin{tabularx}{\textwidth}{CCCCC}
\toprule
\boldmath{$n_0$ [fm$^{-3}$]}&\boldmath{$B_0$ \textbf{[MeV]}}&\boldmath{$K_0$ \textbf{[MeV]}}&\boldmath{$S_0$ \textbf{[MeV]}}\\
\midrule
0.16 & 16 & 240 & 31 \\
\bottomrule
\end{tabularx}
\end{table}
\unskip
\begin{table}[H] 
\caption{Values of meson masses and pion decay constant in the vacuum in unit of MeV.\label{vacM}}
\newcolumntype{C}{>{\centering\arraybackslash}X}
\begin{tabularx}{\textwidth}{CCCCCC}
\toprule
\boldmath{$m_\pi$}&
\boldmath{$m_\eta$}&
\boldmath{$m_{a_0}$}&
\boldmath{$m_\omega$}&
\boldmath{$m_\rho$}&
\boldmath{$f_\pi$}\\
\midrule
138 & 550 & 980 & 783 & 776 & 92.4\\
\bottomrule
\end{tabularx}
\end{table}
\begin{table}[H] 
\caption{Values of $g_1, g_2, \bar{\mu}^2_{\sigma}, \bar{\mu}^2_{a}, \lambda_4,\gamma_4, \lambda_6, g_{\omega NN}$ for $m_0$ = $600-900$ MeV.}
\label{PD1215}
\newcolumntype{C}{>{\centering\arraybackslash}X}
\begin{tabularx}{\textwidth}{CCCCC}
\toprule
\textbf{Parameter} & \textbf{600 MeV} & \textbf{700 MeV} & \textbf{800 MeV} & \textbf{900 MeV}\\
\midrule
$g_1$ & 8.48 & 7.81 & 6.99 & 5.96\\
$g_2$ & 14.93 & 14.26 & 13.44 & 12.41\\
$\bar{\mu}^2_{\sigma}/f^2_{\pi}$ & 22.43 & 19.38 & 12.06 & 1.64\\
$\lambda_4$ & 40.40 & 35.51 & 23.21 & 4.56\\
$\lambda_6 f^2_{\pi}$ & 15.75 & 13.90 & 8.93 & 0.69\\
$g_{\omega NN}$ & 9.14 & 7.31 & 5.66 & 3.52\\
$\bar{\mu}^2_a/f^2_{\pi}$ & -10.77 & -13.82 & -21.15 & -31.56\\
$\gamma_4$ & 180.45 & 168.18 & 135.97 & 84.38\\
\bottomrule
\end{tabularx}
\end{table}

\begin{table}[H] 
\caption{Values of $g_{\rho NN}$ for various $m_0$, $L_0$.}
\label{tab:gr_lor_215}
\newcolumntype{C}{>{\centering\arraybackslash}X}
\begin{tabularx}{\textwidth}{CCCCC}
\toprule
\textbf{$L_0$ [MeV]} & \textbf{600 MeV} & \textbf{700 MeV} & \textbf{800 MeV} & \textbf{900 MeV}\\
\midrule
$L_0=40$ MeV & 15.69 & 14.00 & 12.71 & 11.42\\
$L_0=50$ MeV & 15.20 & 13.46 & 12.07 & 10.71\\
$L_0=60$ MeV & 14.75 & 12.98 & 11.51 & 10.11\\
$L_0=70$ MeV & 14.34 & 12.54 & 11.03 & 9.61\\
$L_0=80$ MeV & 13.96 & 12.15 & 10.60 & 9.17\\
\bottomrule
\end{tabularx}
\end{table}
\unskip
\begin{table}[H] 
\caption{Values of $\lambda_{\omega\rho}$ for various $m_0$, $L_0$.}
\label{tab:lwr_lor_215}
\newcolumntype{C}{>{\centering\arraybackslash}X}
\begin{tabularx}{\textwidth}{CCCCC}
\toprule
\textbf{$L_0$ [MeV]} & \textbf{600 MeV} & \textbf{700 MeV} & \textbf{800 MeV} & \textbf{900 MeV}\\
\midrule
$L_0=40$ MeV & 0.025 & 0.076 & 0.290 & 2.457\\
$L_0=50$ MeV & 0.022 & 0.065 & 0.241 & 1.944\\
$L_0=60$ MeV & 0.019 & 0.054 & 0.192 & 1.430\\
$L_0=70$ MeV & 0.016 & 0.043 & 0.143 & 0.917\\
$L_0=80$ MeV & 0.014 & 0.032 & 0.093 & 0.403\\
\bottomrule
\end{tabularx}
\end{table}
%

\section{Asymmetric nuclear matter properties}
\label{sec:a0}
Neutron star is a highly asymmetric matter 
mainly composed of neutron. Therefore, the properties of asymmetric matter such as the symmetry energy at the 
saturation $S_0$, the slope parameter $L_0$, the symmetry incompressibility $K_{sym}$, the symmetry skewness coefficient $Q_{sym}$, which determine the EoS of the asymmetric matter, have strong impact to the neutron star properties such as their mass and radius. In this section, we compute $K_{sym}$ and $Q_{sym}$. By comparing with the recent constraint of $K_{sym}$, we constrain the chiral invariant mass of nucleon and see whether the constraints for asymmetric nuclear matter properties agree with the NS observations in the present model.

The symmetry energy at arbitrary baryon density is defined as
\begin{equation}
  S(n_B) \equiv \frac{1}{2} \frac{\partial^2 w(x,\delta)}{\partial \delta^2} \Bigg\vert_{\delta = 0 } \ ,
  \label{Sbeq}
\end{equation}
where $w(x,\delta) \equiv \frac{\epsilon (n_B,n_I)}{n_B} - m_N$ is the energy per nucleon with
$x \equiv \frac{n_B - n_0}{3 n_0}$, $\delta \equiv - \frac{2n_I}{n_B}.$
$K_{sym}$ and $Q_{sym}$ are defined as the coefficients of the Taylor expansion of the symmetry energy $S(n_B)$ around the saturation density $n_0$:
\begin{align}
        S(n_B) = {} & S_0 + \left(\frac{n_B - n_0}{n_0}\right)  \frac{L_0}{3} + \left(\frac{n_B - n_0}{ n_0}\right)^2 \frac{K_{sym}}{18}  
        + \left(\frac{n_B - n_0}{n_0}\right)^3  \frac{Q_{sym}}{162} + O(n_B^4) \ , 
\end{align}
where
\begin{align}
    K_{sym} & = 9 n_0^2 \frac{\partial^2 S}{\partial n_B^2} \Biggr|_{n_0} \ , \quad Q_{sym} = 27 n_0^3 \frac{\partial^3 S}{\partial n_B^3} \Biggr|_{n_0}  \ .
\end{align}
They are the higher order coefficients that control the high density behavior of the asymmetric nuclear matter EoS. $K_{sym}$ characterizes the curvature of the symmetry energy with respect to density, analogous to the role of the incompressibility coefficient $K_0$ in symmetric nuclear matter. $Q_{sym}$ encodes how rapidly the curvature of the symmetry energy changes as density increases.

\begin{figure}[h]
\centering
\includegraphics[width=0.5\textwidth]{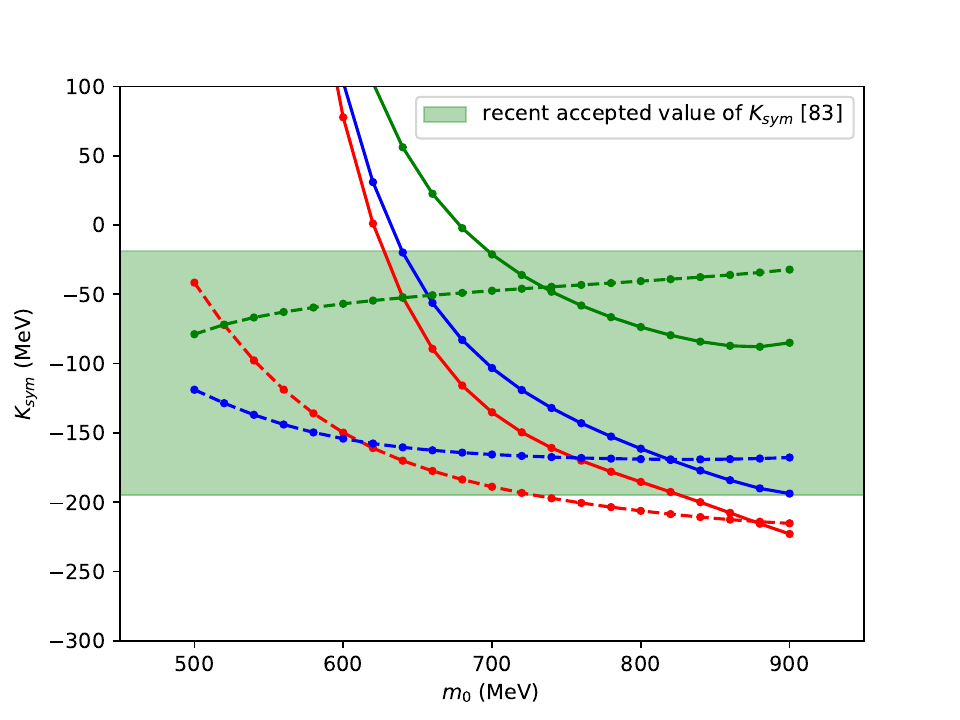} 
\caption{$m_0$ dependence of $K_{sym}$ with $L_0 = 40,57.7,80$ MeV. 
Solid  curves represent the results from
the model with $a_0$ meson and dashed curves are from the model without $a_0$ meson.
The pink region shows
the recent accepted value of $K_{sym}$ as summarized in Ref.~\cite{universe7060182}.
}
\label{m0Ksym}
\end{figure}

Figure~\ref{m0Ksym} shows the $K_{sym}$ as a function of $m_0$ in the models with and without the $a_0$ meson, for various values of $L_0$. The recently accepted value of $K_{sym} = -107 \pm 88$ MeV, as given in Ref.~\cite{universe7060182}, is indicated by the pink band. We observe that the inclusion of the $a_0$ meson has a significant impact on $K_{sym}$, especially when $m_0$ is small. 
In particular, $K_{sym}$ becomes positive and increases rapidly as $m_0$ decreases
in the $a_0$ model:
$K_{sym}>1000$ MeV when $m_0 \lesssim 600$ MeV for $L_0 = 57.7$ MeV. Comparing our results with the recent constraint, we find that the present model imposes a strong constraint on $m_0$, favoring the range $640 \lesssim m_0 \lesssim 860$ MeV for $L_0 = 57.7$ MeV. 
We also note that while larger $L_0$ leads to larger values of $K_{sym}$, 
its influence is weaker than that of $m_0$ 
when $m_0 \lesssim 600$ - $700$\,MeV in the $a_0$ model. However, for larger $m_0$, where the sensitivity to $m_0$ diminishes, the effect of $L_0$ becomes dominant. In contrast, $K_{sym}$ shows much less variation with $m_0$ in the absence of the $a_0$ meson.

\begin{figure}[htb]
\centering
\includegraphics[width=0.5\textwidth]{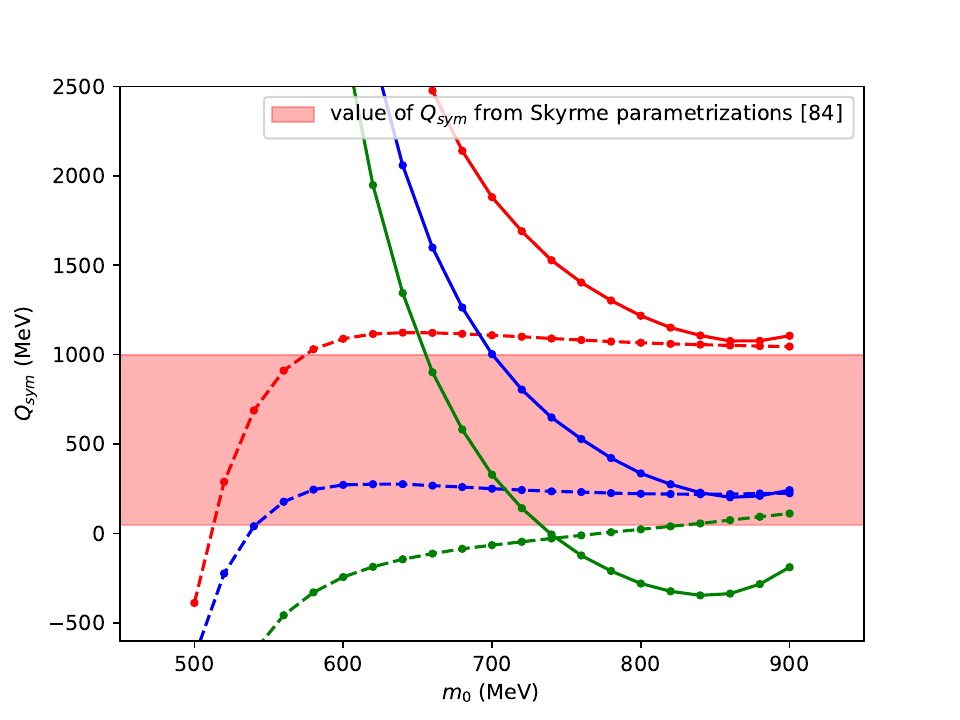} 
\caption{
$m_0$ dependence of $Q_{sym}$ with different $L_0$. Solid curve represents the model with $a_0$ meson and dashed curve represents the model without $a_0$. The blue region is the recent accepted value of $Q_{sym}$ as summarized in Ref.~\cite{PhysRevC.85.035201}.
}
\label{m0Qsym}
\end{figure}

Figure~\ref{m0Qsym} shows the dependence of $Q_{sym}$ on $m_0$ in the present models with and without $a_0$ meson for different $L_0$. The blue band represents the range of $Q_{sym}$ estimated from Skyrme models as summarized in Ref.~\cite{PhysRevC.85.035201}. Similar to the case of $K_{sym}$, the $a_0$ meson have significant impact on $Q_{sym}$. $Q_{sym}$ has a very different $m_0$-dependence in the models with and without $a_0$ meson. While $Q_{sym}$ is negative for small $m_0$ in the model without $a_0$ meson, $Q_{sym}$ is positive and considerably large when $m_0$ is small in the $a_0$ model. In particular, $Q_{sym}$ exceeds several thousand MeV in the $a_0$ model when $m_0 \lesssim 600$ - $700$\,MeV depending on the value of $L_0$, which is much larger than the typical predictions from Skyrme models.

Our results suggest that higher-order asymmetry properties, such as $K_{sym}$ and $Q_{sym}$, are sensitive to the existence of the $a_0$ meson. Notably, the predictions of $K_{sym}$ and $Q_{sym}$ from the present models with $L_0 = 57.7$\,MeV are consistent with recently accepted values and predictions from Skyrme models when $m_0 \approx 700$ - $800$\,MeV. Although $Q_{sym}$ is extremely difficult to be measured experimentally, we believe that future constraints on this quantity could provide valuable insight into the properties of asymmetric nuclear matter, such as the equation of state (EoS) of neutron star matter, and help further our understanding of the chiral invariant mass of the nucleon.

\section{Neutron star matter}
\label{sec:NS}
Neutron stars (NSs) provide unique cosmic laboratories for studying matter under extreme conditions. Recent precise measurements of NS masses and radii have significantly constrained the equation of state (EoS) of strongly interacting matter at densities beyond nuclear saturation\cite{LIGOScientific:2017vwq,LIGOScientific:2018cki,Abbott_2019,Miller:2019cac,Riley:2019yda}. Of particular significance is the discovery of the central compact object in the supernova remnant HESS J1731-347, characterized by an unusually low mass of approximately $0.77^{+0.20}_{-0.17} M_\odot$ and a radius of about $10.4^{+0.86}_{-0.78}$ km \cite{HESSJ1731-347}. This object, being the lightest neutron star ever observed, presents a new challenge for theoretical models that must now accommodate both massive neutron stars ($\sim 2 M_\odot$) and this remarkably light compact object. After including the $a_0$ meson effect, we will study 
the implication of recent NS observations to the nucleon chiral invariant mass.

\subsection{Unified EoS with Crossover Phase Transition}

At densities several times
of nuclear saturation density 
($n_0 \approx 0.16$ fm$^{-3}$), the interior of NS likely undergoes a transition from hadronic to quark degrees of freedom. Traditional approaches often model this as a sharp first-order phase transition, producing discontinuities in the EoS. However, these treatments typically rely on extrapolating hadronic and quark models far beyond their regions of established validity, leading to significant uncertainties in the predicted phase transition behavior and neutron star properties.

In our approach, we adopt a more physically motivated picture of hadron-quark continuity, where the transition occurs smoothly over a finite density range. To implement this, following Refs.~\cite{PhysRevC.103.045205,PhysRevC.106.065205,Minamikawa:2023eky, PhysRevC.109.065807} we expand the pressure as a function of baryon
chemical potential in polynomial form $P(\mu_B)=\sum_{i=0}^5C_i \mu_B^i$.
By imposing six boundary conditions, we interpolate the EoS of
the PDM and that of an NJL-type quark model as constructed in Ref.~\cite{PhysRevC.103.045205} in the intermediate density region $2n_0\leq n_B\leq 5n_0$, to obtain a smoothly
connected unified EOS.

\subsection{NS mass-radius relation}
By solving the Tolman-Oppenheimer-Volkoff (TOV) equation for spherically symmetric and static stars, we obtain the NS mass-radius ($M$-$R$) relation.

\begin{figure}[htb]
\centering
\includegraphics[width=0.5\textwidth]{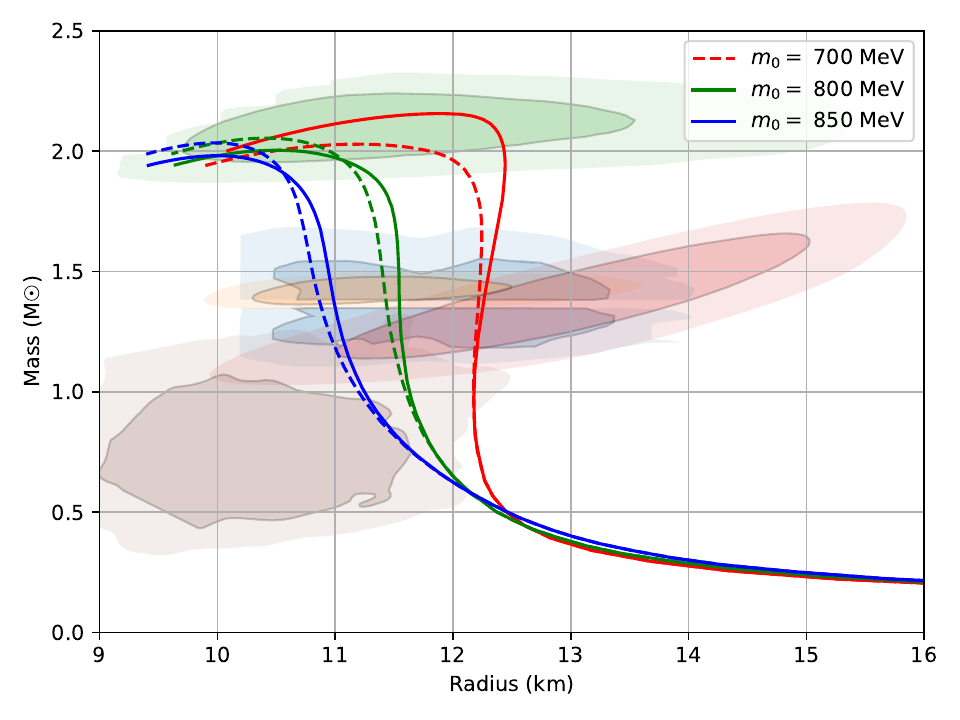}
\caption{
Mass-radius relation for $m_0=700, 800, 850$ MeV with fixed value of $L=40$ MeV connected with different combination of NJL model parameter $H$ and $g_V$. Blue curve is connected with $(H, g_V)/G=(1.5, 0.7), (1.55, 0.8)$; green curve is connected with $(H, g_V)/G=(1.45, 0.7), (1.5, 0.8)$; red curve is connected with $(H, g_V)/G=(1.4, 0.7), (1.4, 0.8)$. See Ref.~\cite{PhysRevC.103.045205} for details of the NJL model.
}
\label{mr_eg}
\end{figure}

\begin{figure}[h]
\centering
\includegraphics[width=0.5\textwidth]{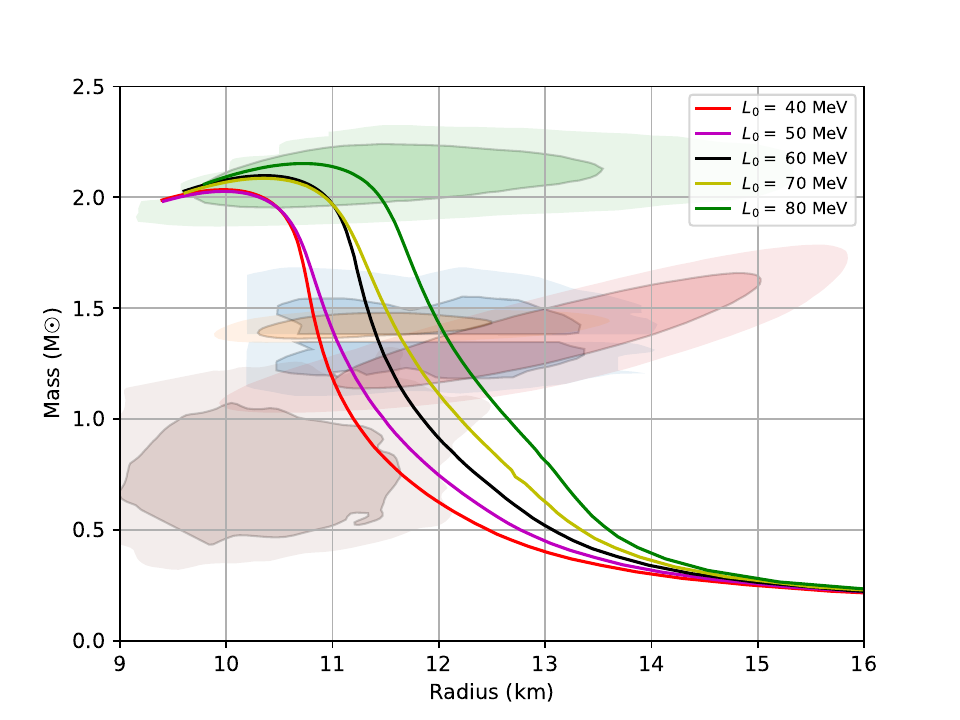} 
\caption{
M-R relations for $m_0=$ 850 MeV with different $L_0$. The red curve is connected to ($H/G$,$g_v/G$) = (1.55,0.8); the purple curve is connected to ($H/G$,$g_v/G$) = (1.55,0.8); the black curve is connected to ($H/G$,$g_v/G$) = (1.55,0.9); the yellow green curve is connected to ($H/G$,$g_v/G$) = (1.55,0.9); the green curve is connected to ($H/G$,$g_v/G$) = (1.55,1). See Ref.~\cite{PhysRevC.103.045205} for details of the NJL model.
}
\label{fig:changeL}
\end{figure}

In the PDM, the chiral invariant mass $m_0$ and the slope parameter $L_0$ play crucial roles in determining the stiffness of the EoS. Larger values of $m_0$ lead to a softer EoS in the hadronic region, while larger values of the slope parameter $L_0$ result in a stiffer EoS. 
To illustrate these effects, we examine the NS $M$--$R$ relations under different parameter combinations. In Fig.~\ref{mr_eg}, we fix $L_0 = 40$ MeV and vary the chiral invariant mass from $m_0 = 700$ MeV to $850$ MeV. As $m_0$ increases, the hadronic EOS becomes progressively softer, resulting in $M$--$R$ curves with systematically smaller radii for any given mass. Conversely, in Fig.~\ref{fig:changeL}, we fix $m_0 = 850$ MeV and vary $L_0$ from $40$ to $80$ MeV. Here, decreasing $L_0$ values correspond to softer EoS and smaller radii. By treating these two parameters as variables, we can investigate how recent neutron star observations constrain their allowed values in the present model.


\begin{figure}[h]
\centering
\includegraphics[width=0.5\textwidth]{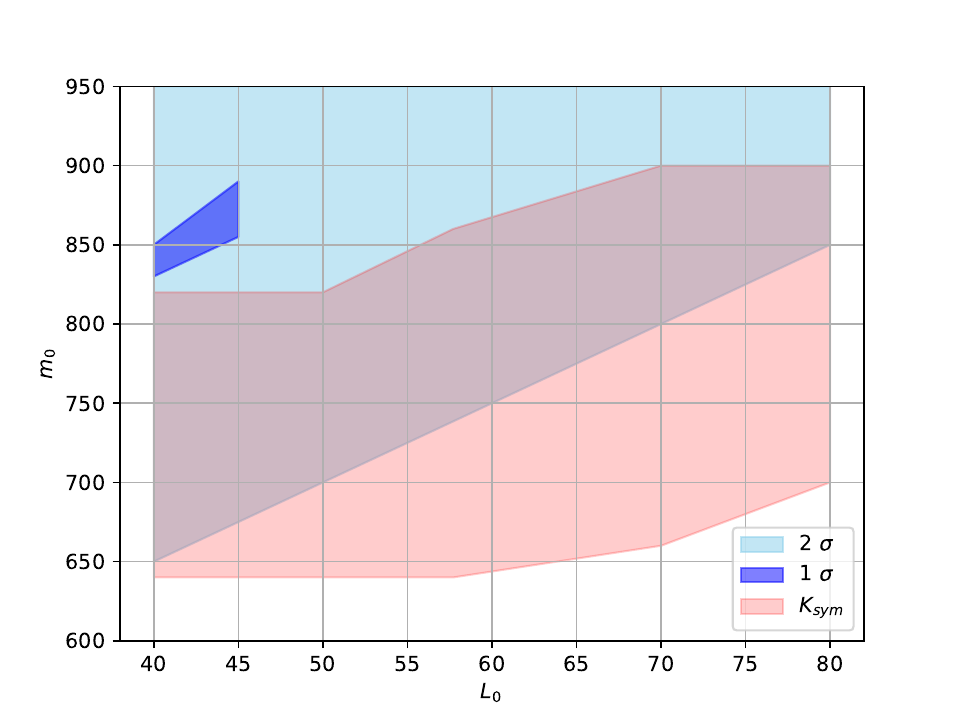} 
\caption{
Allowed region for $m_0$ and $L_0$. The blue region shows the value of $m_0$ and $L_0$ which the MR relations satisfy the 1 $\sigma$ and 2 $\sigma$ constraints from the NSs observational data. The pink region shows the constraint from symmetry incompressibility $K_{sym}$ 
}
\label{m0L0constraint}
\end{figure}

We compare our results with the observational data of HESS J1731-347, PSR J0437–4715, GW170817, PSR J0740+6620, and PSR J0030+0451 to constrain the value of $m_0$ and $L_0$. Figure~\ref{m0L0constraint} shows the constraint to $m_0$ as a function of $L_0$ from the 1 $\sigma$ and 2 $\sigma$ observational constraints of the above NSs. The constraint from the symmetry incompressibility $K_{sym}$ presented in Ref.~\cite{QCS2023KONG} is also included for comparison. 
We observe that the 1$\sigma$ neutron star (NS) constraint is very tight, allowing only a narrow region in the $(m_0, L_0)$ plane to satisfy the condition, as indicated by the dark-blue band in Fig.~\ref{m0L0constraint}. The 1$\sigma$ NS constraint appears to be in slight tension with the constraint from $K_{sym}$, although the dark-blue band lies close to the $K_{sym}$-derived region. This discrepancy may arise from the uncertainties inherent in both theoretical models and experimental extractions of $K_{sym}$, reflecting the challenges in determining higher-order symmetry energy coefficients such as $L_0$ and $K_{sym}$. Nevertheless, there is an overlap region between the 2 $\sigma$ NS constraint and the $K_{sym}$ constraint,
which restricts the allowed range of $m_0$ to
\begin{equation}
   740 \,\text{ MeV} \lesssim m_0 \lesssim 860 \,\text{ MeV} \ ,
\end{equation} 
for $L_0=57.7$\,MeV.
Compared with previous results in Ref.~\cite{PhysRevC.109.065807}, the constrained $m_0$ is shifted to larger values due to the stiffening effects from the $a_0$ meson.
In this work, we do not consider the constraint from $Q_{sym}$ because it is not well-determined experimentally. Future experiments on the asymmetric matter EoS will help us to further constrain the chiral invairant mass of nucleon as well as the behavior of asymmetric matter at high density.

\section{Summary}\label{sec:summary}

In this work, we first studied the effect of the $a_0$ meson to the higer order asymmetric matter properties such as the symmetry incompressibility $K_{sym}$ and the symmetry skewness $Q_{sym}$. We find that $K_{sym}$ and $Q_{sym}$ is sensitive to the chiral invariant mass of nucleon $m_0$ in the presence of $a_0$ meson.

Then, we studied the neutron star $M$-$R$ relation and gave constraints to the slope parameter $L_0$ and $m_0$. The ultra-light compact object HESS J1731-347 provides particularly stringent constraints on our model parameters. With its unusually small radius and low mass, this object requires a very soft EoS in the hadronic region, which requires  large $m_0$ and small $L_0$ values in our model. Our calculations demonstrate that with 
$740 \,\text{ MeV} \lesssim m_0 \lesssim 860 \,\text{ MeV}$ for $L_0 = $ 57.7 MeV, 
our unified EoS can simultaneously satisfy all observational constraints within 2 $\sigma$ credible region including the HESS J1731-347 observation, as well as the constraint from asymmetric nuclear matter properties such as $K_{sym}$. On the other hand, the 1$\sigma$ data from HESS J1731–347 impose a very narrow constraint on the allowed values of $m_0$ and $L_0$. In addition, the 1$\sigma$ constraint from neutron stars appears to be not fully compatible with the constraint from $K_{sym}$. This discrepancy may arise from the uncertainties in determining $K_{sym}$ and the radius of HESS J1731-347. Nevertheless, the constraints from neutron stars show good overall agreement with the $K_{sym}$ constraint within the 2$\sigma$ level. This finding suggests that, if confirmed as a neutron star, HESS J1731-347 would significantly narrow the allowed parameter space of the model, offering valuable insights into the nature of the chiral invariant mass of nucleon and the behavior of dense asymmetric matter that are difficult to probe in terrestrial experiments.

\vspace{6pt}

\authorcontributions{Writing—original draft preparation, Y.-K.K., B.G., and M.H.; writing—review and editing, Y.-K.K., B.G., and M.H. All authors have read and agreed to the published version of the manuscript. 
}

\funding{
This work is supported in part by JSPS KAKENHI Grants,No. 23H05439, and No. 24K07045 and JST SPRING,
Grant No. JPMJSP2125. B.G. would like to take this opportunity to thank the “Interdisciplinary Frontier Next-Generation
Researcher Program of the Tokai Higher Education and Research System.”
}

\dataavailability{Data is contained within the article. } 

\acknowledgments{The authors would like to thank the organizers of ``Compact Stars in the QCD phase diagram (CSQCD2024)'' for giving this opportunity to write the contribution.} 

\conflictsofinterest{The authors declare no conflicts of interest. 
}

\begin{adjustwidth}{-\extralength}{0cm}

\reftitle{References}


\bibliography{PDM-a0-refs}

\PublishersNote{}
\end{adjustwidth}
\end{document}